\documentclass[12pt]{article}
\usepackage[export]{adjustbox}
\usepackage{graphicx}
\usepackage{graphics}
\usepackage{epsfig}
\usepackage{amssymb}
\usepackage{amsmath}
\usepackage{color}
\usepackage{textcomp}
\usepackage{latexsym}
\usepackage{physics}

\usepackage[margin=2.5cm,footskip=0.6cm]{geometry}
\def\bea{\begin{eqnarray}}
\def\eea{\end{eqnarray}}
\def\beq{\begin{equation}}
\def\eeq{\end{equation}}

\def\bm{\begin{math}}
\def\me{\end{math}}

\begin{document}

\begin{center}
{\Large{\bf  Domain Growth Kinetics in Active Binary Mixtures}}\\
\ \\
\ \\
by \\
Sayantan Mondal and Prasenjit Das\footnote{prasenjit.das@iisermohali.ac.in}  \\
Department of Physical Sciences, Indian Institute of Science Education and Research Mohali, Knowledge City, Sector 81, SAS Nagar, Mohali, Punjab, 140306, India\\
\end{center}

\begin{abstract}
\noindent We study motility-induced phase separation (MIPS) in symmetric and asymmetric active binary mixtures. We start with the coarse-grained run-and-tumble bacterial model that provides evolution equations for the density fields $\rho_i(\vec r, t)$. Next, we study the phase separation dynamics by solving the evolution equations using the Euler discretization technique. We characterize the morphology of domains by calculating the equal-time correlation function $C(r, t)$ and the structure factor $S(k, t)$, both of which show dynamical scaling. The form of the scaling functions depends on the mixture composition and the relative activity of the species, $\Delta$. For $k\rightarrow\infty$, $S(k, t)$ follows Porod's law: $S(k, t)\sim k^{-(d+1)}$ and the average domain size $L(t)$ shows a diffusive growth as $L(t)\sim t^{1/3}$ for all mixtures.
\end{abstract}

\newpage
\section{Introduction}
\label{sec1}
An active matter system is a collection of self-propelled entities, often referred to as ``active agents" or ``active particles," that can convert stored or ambient energy into directed motion. The typical size of such systems ranges from a few $\mu$m (e.g., actin and tubulin filaments, molecular motors, sperm cells, microorganisms such as amoeba and bacteria~\cite{NSML97,HNKY8786,AIJR007}) to several meters (e.g., schools of fish~\cite{MW98}, bird flocks~\cite{DYGMYW19}, human crowds). These systems are always out of equilibrium due to the intrinsic activity of their constituents~\cite{MJSTJR13,S10}, resulting in a variety of collective phenomena and emergent features that are not observed in equilibrium or passive systems. Active phase separation~\cite{SRJ15,JCAMDAW12,AKTP21}, active turbulence~\cite{ActTur}, active crystallization~\cite{ActCry}, and the development of complex patterns such as vortices, bands~\cite{AJHJ15}, and so on are some examples. Understanding and characterizing such collective behaviors and dynamics in active matter systems  has garnered inspiration from a variety of disciplines over the last two decades, including physics, biology, chemistry, and engineering.  One of the key features of active matter is the emergence of motility-induced phase separation (MIPS)~\cite{JS21,OSTZ21} and pattern formation~\cite{MDIJ10}. Active particles can self-organize into dynamic structures such as swarms~\cite{JYS05,TAEIO95}, flocks, vortices, and clusters. These patterns often result from the interplay between self-propulsion, particle-particle interactions, and the influence of the surrounding medium.

Over the last several decades, a great deal of research has been done on the kinetics of phase separation in passive A$_{1-y}$B$_y$ binary mixtures~\cite{A94}. The Cahn-Hilliard-Cook (CHC)~\cite{H70} equation successfully describes the diffusion-driven phase separation in passive binary mixtures, also referred to as Model B~\cite{JJ58,HH77}, is given by
\begin{eqnarray}
\label{neweqn1}
\frac{\partial}{\partial t}\psi(\vec r, t) = \vec\nabla\cdot\left[D_p\vec\nabla\left( \frac{\delta F[\psi]}{\delta\psi}\right) + \vec\xi(\vec r, t)\right].
\end{eqnarray}
Here, $D_p$ is the diffusion coefficient and $\psi(\vec r, t)$ is the order-parameter. The Helmholtz potential $F[\psi]$ has the standard $\psi^4$ form~\cite{DP04,PW09}:
\begin{eqnarray}
\label{neweqn2}
F[\psi]=\int d\vec r\left[-\frac{a(T_c - T)}{2}\psi^2 + \frac{b}{4}\psi^4 + \frac{\zeta_p}{2} \left(\vec\nabla\psi\right)^2 \right],
\end{eqnarray}
where the parameters $a,b,\zeta_p>0$. The vector Gaussian white noise $\vec\xi(\vec r, t)$ satisfies the usual fluctuation-dissipation relation:
\begin{eqnarray}
\label{neweqn3}
\langle\vec\xi(\vec r, t)\rangle = 0 ~~~\text{and}~~~ 
\langle\xi_\mu(\vec r_1, t_1)\xi_\nu(\vec r_2, t_2)\rangle = 2D_pk_{\rm B}T\delta_{\mu\nu}\delta(\vec r_1 - \vec r_2)\delta(t_1 - t_2).
\end{eqnarray}
Here, $\mu$ and $\nu$ refer to Cartesian coordinates. For a symmetric binary mixture ($y=0.5$), i.e., a mixture with an equal amount of A and B, domains are interconnected, while we see the circular droplet morphology of the minority component for asymmetric mixtures ($y\ne0.5$), where the amount of A and B varies. The morphology of domain growth has been studied using the equal-time correlation function and structure factor, which show dynamical scaling with the scaling forms~\cite{DP04,PW09}:
\begin{eqnarray}
\label{neweqn4}
C(\vec r, t)  = g_p\left[r/L(t)\right] ~~~\text{and}~~~ 
S(\vec k, t)  = L(t)^df_p\left[kL(t)\right] 
\end{eqnarray}
Here, $g_p(x)$ and $f_p(q)$ are time independent master functions that depend on $y$~\cite{SP88}, and $d$ is the spatial dimensionality. Notably, these functions are independent of $T$ or the noise amplitude in the scaling regime, as defined by eq.~(\ref{neweqn3}). The time-dependent average domain size $L(t)$ shows a power-law growth: $L(t) = A_pt^{1/3}$ which is known as Lifshitz-Slyozov~(LS) law~\cite{LS61}. The prefactor $A_p$ depends on $D_p$ and other system-specific factors.

Understanding phase separation in active mixtures~\cite{M12,MJ15} has gained enormous attention in recent years. Active particles' motion is not solely determined by external forces or thermal fluctuations, unlike passive particles. As a result, active systems break the detailed balance criterion~\cite{JSAVP17} and do not meet the fluctuation-dissipation theorem, resulting in a drastic effect on the kinetics of phase ordering. Before going into the details of MIPS in binary active mixtures, let us briefly summarize some of the earlier experimental and theoretical works on MIPS in one-component systems. The latter shows phase separation between dense and dilute phases. The first experimental study of MIPS is due to Dombrowski {\it et al}~\cite{CLSRJ004}. They observed large-scale coherent structure formation in bacterial dynamics in a confined geometry. Next, experimental studies on synthetic active particles; e.g., Janus particles~\cite{ICJCL12,FIFCC18} show various self-assemblies those are tunable by controlling environment and confinement. Buttinoni {\it et al.}~\cite{IJFJCT13} studied dynamical clustering and phase separation in suspensions of self-propelled colloidal particles experimentally as well as by computer simulation. Geyer {\it et al.}~\cite{DDJD19} studied motility-induced freezing transitions in polar active liquids. Recently, Anderson and Fernandez-Nieves studied social interaction-mediated motility-induced phase separation in fire ants~\cite{CA22}.  The first coarse-grained study of MIPS is due to Tailleur and Cates~\cite{JM08}. Starting with the microscopic dynamics of a single particle, they have obtained an evolution equation for the density of a system of self-propelled particles undergoing run-and-tumble dynamics. They observed MIPS between dense and dilute phases when the self-propulsion speed $v$ is a sufficiently rapidly decreasing function of the local density $\rho$. They also discovered that in some situations, the system's free-energy density can be mapped onto systems satisying detailed-balance criterion. Later, they extended their study to the active Brownian particles (ABPs)~\cite{JM13,WSC17}. Stenhammar {\it et al.} studied the phase behavior of an active Brownian particle system using a coarse-grained model~\cite{STMAC2013} and Brownian dynamics simulation~\cite{SMAC2014} in different spatial dimensions, $d$. They found that the domain growth law depends on $d$. Later, Wittkowski {\it et al.}~\cite{RAJRDM14} introduced a phenomenological
 active model B (AMB) to study phase separation dynamics in a one-component active system that undergoes liquid-vapour phase transition. They modeled the detailed balance violation by introducing a lowest order gradient term $\alpha|\vec\nabla\psi|^2$ in the evolution equation of Model B. By tuning $\alpha$, they controlled the activity of the system ($\alpha=0$ corresponds to standard Model B). The average domain size $L(t)$ shows a power-law growth $L(t)\sim t^\phi$ with an exponent $\phi$ that depends on $\alpha$.

There are experimental and theoretical studies on phase separation in active binary mixtures. The first experimental study on the phase separation of bacterial mixtures interacting via Quorum sensing is due to Curatolo {\it et al.}~\cite{ANYCAJJ20}. In this work, two strains of E. coli bacteria were engineered to cross-regulate each other’s motility and left to grow in a Petri dish. Depending on the type of interaction, whether it was mutual inhibition or mutual activation of motility, two different emergent patterns were observed, namely mixed or demixed concentric rings. In recent times, Zheng {\it et al.}~\cite{JJYYXJ23} studied the photo-induced phase separation in binary colloidal mixtures. In their experiment, the average domain size shows a power law growth with an exponent that depends on photo intensity. Recently, Pattanayak {\it et al.}~\cite{SSS21,SSS2} extended the phenomenological model of Wittkowski {\it et al.}~\cite{RAJRDM14} (AMB) for A$_{1-y}$B$_y$ active binary mixtures and provided a detailed study for symmetric and asymmetric mixtures. For any negative non-zero $\alpha$, they observed circular droplet domain morphology for a symmetric mixture and the domain growth exponent shows a crossover from 1/3 at early times to 1/4 at a later time. For a given $\alpha$, the domain morphology is statistically self-similar and characterized by the equal-time correlation function $C(\vec r, t)$ and structure factor $S(\vec k, t)$, which exhibit dynamical scaling of the form
\begin{eqnarray}
\label{apeqn1}
C(\vec r, t)  = g_a^\alpha \left[r/L(t)\right] ~~~\text{and}~~~ 
S(\vec k, t)  = L(t)^df_a^\alpha \left[kL(t)\right] 
\end{eqnarray}
Here $g_a^\alpha(x)$ and $f_a^\alpha(q)$ are the $\alpha$-dependent master functions, which characterize the domain morphology. Additionally, they have studied the effects of additive and multiplicative noises on the domain morphologies in symmetric and asymmetric mixtures~\cite{SSS2}. In their model, the noise-noise correlation is independent of $\alpha$. They found that the presence of noise did not lead to any significant differences in the domain morphologies. Saha {\it et al.}~\cite{SJR20} proposed a continuum model of pattern formation by considering a nonreciprocal interaction between multiple species of scalar active matter. They modeled the nonreciprocity by modifying the chemical potential. Concurrently, You {\it et al.}~\cite{ZAM20} studied a similar model by introducing nonreciprocity via cross-diffusivities. They observed bulk phase separation, traveling waves, and oscillatory phases in a phase-separating binary mixture depending on the nonreciprocity parameter and composition. However, a detailed study of domain growth kinetics in MIPS using a coarse-grained model derived from run-and-tumble bacterial motion is absent from all previous research.

In this paper, we provide a comprehensive study of MIPS in symmetric and asymmetric active binary mixtures in $d=2$. We employ the coarse-grained dynamical model of run-and-tumble particles proposed by Curatolo {\it et al}~\cite{ANYCAJJ20,CuraT}. The dynamical equations resemble the CHC equation~(\ref{neweqn1}) in form. Our primary objective is to study the asymptotic domain growth laws and the dynamical scaling of domain morphologies, comparing them with Model B and AMB.

The organization of the paper is as follows: In Section 2, we have described the model. Numerical details and results are described in Section 3. Finally, we conclude our paper in Section 4.

\section{Details of Modeling}
\label{sec2}
We begin with a two dimensional run and tumble bacterial model with tumbling rate $\beta$ and a spatially varying speed $v(\vec r)$~\cite{JM08,JM13}. The corresponding master equation for joint probability density $P(\vec r,\theta,t)$ is following
\begin{eqnarray}
\label{apeqn3}
\frac{\partial P(\vec r, \theta, t)}{\partial t} = -\vec\nabla \cdot \left[v(\vec r)\vec u(\theta) P(\vec r, \theta, t)\right] - \beta P(\vec r, \theta, t) + \frac{\beta}{2\pi}\int d\theta^{\,\prime} P(\vec r, \theta^{\,\prime}, t).
\end{eqnarray}
Here $\vec r$ is the position, $\theta$ is the polar angle and $\vec u$ is the unit vector representing the particle's orientation. Next, we define $Q(\vec r,t)$ as the probability density to find the particle at position $\vec r$ irrespective of it's orientation as
\begin{eqnarray}
    Q(\vec r,t) = \int d\theta P(\vec r, \theta, t)
\end{eqnarray}
To construct the long-time large-scale dynamical behavior of $Q(\vec r,t)$, we integrate eq.~(\ref{apeqn3}) over $\theta$ to get
\begin{eqnarray}
    \label{Neq5}
    \frac{\partial Q(\vec r, t)}{\partial t} = -\vec\nabla \cdot \left[v(\vec r)\vec m(\vec r, t)\right].
\end{eqnarray}
Here $\vec m(\vec r, t) = \int d\theta \vec u(\theta) P(\vec r, \theta, t)$ is the average orientation of the particle. Equation~(\ref{Neq5}) shows that time evolution of $Q(\vec r,t)$ involves $\vec m(\vec r, t)$, which is the 1st order moment of $\vec u$. It has been shown that the evolution equation of $\vec m(\vec r, t)$ involves higher-order moments of $\vec u$ and so on~\cite{newbook}. For large system sizes $\mathcal{L}\rightarrow\infty$, $m(\vec r, t)$ and its higher-order moments relax at a time scale $\beta^{-1}$, so they are fast variables~\cite{JS21,OSTZ21}. Using this information, one can obtain closure relations for the higher-order moments of $\vec u$ and derive the large-scale long-time dynamics of $Q(\vec r,t)$ as
\begin{eqnarray}
\label{apeqn4}
\frac{\partial Q(\vec r,t)}{\partial t} &= \vec \nabla \cdot \left[ D(\vec r)\vec \nabla Q-\vec F(\vec r)Q\right],
\end{eqnarray}
with diffusivity $D(\vec r)=v(\vec r)^2/2\beta$ and effective force $\vec F(\vec r)=-v(\vec r)\vec \nabla v(\vec r)/2\beta$. The eq.~(\ref{apeqn4}) is equivalent to an Ito-Langevin dynamics~\cite{JM08} as
\begin{eqnarray}
    \label{Neq6}
    \dot{\vec r} = \vec F(\vec r) + \sqrt{2D(\vec r)}\vec \eta(t),
\end{eqnarray}
where $\vec \eta(t)$ is a vector Gaussian white noise satisfying $\langle\eta_\mu(t_1)\eta_\nu(t_2) \rangle=\delta_{\mu\nu}\delta(t_1 - t_2)$.

Next, we construct the fluctuating hydrodynamics for $N$ non-interacting active particles. Using Ito calculus, it has been shown that the stochastic evolution of the density field $\rho(\vec r,t)$ obeys the following dynamical equation~\cite{D96}:
\begin{eqnarray}
\label{apeqn5}
\frac{\partial \rho(\vec r,t)}{\partial t} &= \vec\nabla \cdot \left[ D(\rho)\vec \nabla \rho -\vec F(\rho)\rho + \sqrt{2D(\rho)\rho}\vec\Lambda(\vec r, t)\right],
\end{eqnarray}
with diffusivity $D(\rho)=v(\rho)^2/2\beta$, effective force $\vec F(\rho)=-v(\rho)\vec \nabla v(\rho)/2\beta$ and Gaussian white noise $\vec\Lambda(\vec r, t)$ satisfies $\langle\Lambda_\mu(\vec r_1, t_1)\Lambda_\nu(\vec r_2, t_2)\rangle=\delta_{\mu\nu}\delta(\vec r_1 - \vec r_2)\delta(t_1-t_2)$. Further, $\vec F(\rho)$ can be rewritten as
\begin{eqnarray}
\label{apeqn6}
 \vec F(\rho)=-\frac{v(\rho) v'(\rho)}{2\beta}\vec \nabla\rho,
\end{eqnarray}
which reduces eq.~(\ref{apeqn5}) to a diffusion equation analogous to the CHC equation~(\ref{neweqn1}) as
\begin{eqnarray}
\label{apeqn7}
 \frac{\partial \rho(\vec r,t)}{\partial t}&=\vec \nabla \cdot\left[D_{\rm eff}(\rho)\vec\nabla \rho + \sqrt{2D(\rho)\rho}\vec\Lambda(\vec r, t) \right]\quad\text{with}\quad D_{\rm eff}(\rho)=D(\rho)+\rho\frac{D'(\rho)}{2}.
\end{eqnarray}
Here, the prime denotes differentiation with respect to $\rho$. The effective diffusivity $D_{\rm eff}(\rho)$ can be negative when $v(\rho)$ decreases rapidly, leading to the phase separation between two coexisting phases. Moreover, the diffusivity $D(\rho)$ in the amplitude of noise is different from $D_{\rm eff}(\rho)$, indicating the out-of-equilibrium nature of the active system.

Further, Curatolo {\it et al.} showed that the above procedure can be generalized for the $n$-component systems~\cite{ANYCAJJ20,CuraT}. The evolution equation for the  density field of $i$th component in $d$-dimension reads as
 \begin{eqnarray}
\label{apeqn8}
\frac{1}{d\beta}\frac{\partial\rho_{i}(\vec r,t)}{\partial t}=\vec\nabla\cdot\left[v_{i}^2\vec\nabla\rho_{i}(\vec r,t)+v_{i}\rho_{i}(\vec r,t) \vec\nabla v_{i} + \vec\Theta_i(\vec r, t) \right]- \zeta\nabla^4\rho_{i}(r,t).
\end{eqnarray}
Here, $\vec\Theta_i(\vec r, t) = \sqrt{\rho_i/\beta}v_i\vec\Lambda(\vec r, t)$ and $v_i\equiv v_i(\{\rho_{i=1,n}\})$ is the speed of $i$th species. We have phenomenologically added the term $\zeta\nabla^4\rho_i$ to stabilize the interfaces between different phases, with $\zeta$ being the strength of surface tension. This term, however, can be derived from microscopic dynamics by considering a slow spatial variation of coarse-grained density and approximating it through a gradient expansion~\cite{CuraT}. We set $\zeta>0$ to induce phase separation.

Tailleur and Cates have shown that a system with self-propelled particles and quorum sensing can be mapped to an equilibrium system whose dynamics can be obtained from a free energy functional~\cite{JM08}. This approach can be generalized for multi-species systems having $n$-components with local interactions; the free energy density takes the following form~\cite{CuraT}
\begin{eqnarray}
\label{apeqn9}
f(\{\rho_n\}) = \sum_n \rho_n (\ln \rho_n - 1) + f_{\text{ex}}(\{\rho_n\}).
\end{eqnarray}
The first term corresponds the free energy density of an ideal gas, and $f_{\text{ex}}(\{\rho_n\})$ represents the excess free energy density. The Schwarz's theorem shows that $f_{\text{ex}}(\{\rho_n\})$ will exist and moreover will be continuous only if speeds $v_i$ take the form:
\begin{eqnarray}
\label{apeqn10}
v_i(\{\rho_{n \neq i}\}) = v_i^0 \exp\left( \lambda \prod_{n \neq i} \rho_n\right),
\end{eqnarray}
where $\lambda$ is the interaction parameter and $v_i^0$ represents the amplitude of the speed $v_i$. For phase separation in an AB binary mixture, eq.~(\ref{apeqn10}) yields
\begin{eqnarray}
\label{apeqn1011}
v_{A}(\rho_B)=v_A^0\exp(\lambda\rho_B) ~~~\text{and}~~~v_{B}(\rho_A)=v_B^0\exp(\lambda\rho_A).
\end{eqnarray}
The free energy density takes the form 
\begin{eqnarray}
\label{Neqn13}
f(\rho_{\rm A},\rho_{\rm B}) = \rho_{\rm A} (\ln \rho_{\rm A} - 1) + \rho_{\rm B} (\ln \rho_{\rm B} - 1) + \lambda\rho_{\rm A}\rho_{\rm B}.
\end{eqnarray}

We define a Hessian matrix $\mathcal{H}$ as 
\begin{eqnarray}
    \label{hessian}
    \mathcal{H}_{ij}=\frac{\partial^2 f}{\partial\rho_i\partial\rho_j}.
\end{eqnarray}
The concavity condition on $f$ demands that at least one of the eigenvalues of $\mathcal{H}$ must be negative. This means det$(\mathcal{H})<0$, leading to the instability condition for the binary case \begin{eqnarray}
\label{apeqn11}
\lambda^2>\frac{1}{\rho_A\rho_B}.
\end{eqnarray}
The sign of the parameter $\lambda$ determines whether the two species are mutually inhibiting ($\lambda < 0$), which leads to colocalization, or mutually activating ($\lambda > 0$), which leads to phase separation.

Next, non-dimensionalize eq.~(\ref{apeqn8}) for a binary mixture together with eqs.~(\ref{apeqn1011}) and (\ref{apeqn11}) by rescaling
$$v_i = v_0\Tilde{v_i},~~ \vec r = \sqrt{\zeta/(v_0)^2}\Tilde{\vec r},~~ t = \frac{\zeta}{d\beta (v_0)^4}\Tilde{t},~~\rho_i=\frac{\Tilde{\rho_i}}{|\lambda|},~~ \vec\Theta_{i}=\frac{v_0^3}{|\lambda|\sqrt{\zeta}}\tilde{\vec\Theta}_{i}.$$
Here, all the variables with a tilde are dimensionless quantities, $v_0$ is a scale of speed, and $i\in(A,B)$. In terms of rescaled variables (dropping tildes), eq.~(\ref{apeqn8}) reduces to
\begin{eqnarray}
\label{apeqn81}
\frac{\partial\rho_{i}(\vec r,t)}{\partial t}=\vec\nabla\cdot\left[v_{i}^2\vec\nabla\rho_{i}(\vec r,t)+v_{i}\rho_{i}(\vec r,t) \vec\nabla v_{i}+\vec{\Theta_{i}}(\vec{r},t)\right]- \nabla^4\rho_{i}(\vec r,t),
\end{eqnarray}
Here, we set $v_A^0=v_0$ and $v_B^0=v_0\Delta$, where $\Delta$ represents the relative activity of the two species. Consequently, eq.~(\ref{apeqn1011}) changes to
\begin{eqnarray}
\label{apeqn101}
v_{A}(\rho_B)=\exp(\pm\rho_B) ~~~\text{and}~~~v_{B}(\rho_A)=\Delta\exp(\pm\rho_A),
\end{eqnarray}
where the + and - signs correspond to phase separation and colocalization, respectively. The Gaussian white noise $\vec{\Theta_{i}}(\vec{r},t)$ obeys the following relations:
\begin{eqnarray}
\label{apeqn82}
\langle\vec\Theta_i(\vec{r},t)\rangle=0~\text{and}~
\langle\Theta_{i,\mu}(\vec{r}_1, t_1)\Theta_{j,\nu}(\vec{r}_2, t_2)\rangle=\epsilon \rho_i v_i^2\delta_{ij} \delta_{\mu\nu} \delta(\vec{r}_1 - \vec{r}_2) \delta(t_1 - t_2)~\text{with}~\epsilon = \frac{d|\lambda|v_0^d}{\zeta^{\frac{d}{2}}},
\end{eqnarray} 
where $i$ and $j$ stand for species. Finally, the instability condition given by eq.~(\ref{apeqn11}) becomes
\begin{eqnarray}
\label{apeqn111}
\rho_A\rho_B>1.
\end{eqnarray}
\section{Numerical Simulations And Results}\label{sec3}
We numerically solve eq.~(\ref{apeqn81}) using the Euler discretization method in $d=2$ for an A$_{1-y}$B$_y$ active binary mixture. The speed of $i$th species ($i\in A,B$) $v_i$ is obtained by using eq.~(\ref{apeqn101}). Our system size is $L_x\times L_y=512\times 512$ for both symmetric ($y=0.5$) and asymmetric ($y\ne0.5$) mixtures. Table~\ref{Table_1} lists the initial homogeneous densities of species for various values of $y$. We vary $\epsilon$ and $\Delta$ in our numerical simulations across different mixture compositions. For numerical stability, we choose the discretization mesh sizes to be $\Delta x=0.5$ and $\Delta t=0.001$. We define the order parameter as $\psi(\vec r,t)=[\rho_{A}(\vec r,t) - \rho_{B}(\vec r,t)]/[\rho_{A}(\vec r,t) + \rho_{B}(\vec r,t)]$, where $\rho_i(\vec r,t)$ is the local concentration of $i$th species at position $\Vec{r}$ and time $t$. Therefore, regions with $\psi>0$ and $\psi<0$ will, respectively, correspond to A-rich and B-rich domains. We apply periodic boundary conditions in all directions. All the statistical quantities are averaged over 10 independent runs. 
\begin{table}[h!]
\centering 
\caption{Initial homogeneous densities of $A$ and $B$ for different $y$.}
\label{Table_1}
\begin{tabular}{|p{2cm}|p{2cm}|p{2cm}|}
  \hline
  $y$ & $\rho_A^0$ & $\rho_B^0$ \\ \hline
  0.50 & 1.065 & 1.065 \\ \hline
  0.55 & 0.959 & 1.172 \\ \hline
  0.60 & 0.865 & 1.298 \\ \hline
  0.65 & 0.778 & 1.444 \\ \hline
  0.70 & 0.695 & 1.625 \\ \hline 
\end{tabular}
\end{table} 

\subsection{Symmetric Case}\label{sec3a}
\begin{figure}
\centering
\includegraphics*[width=0.60\textwidth]{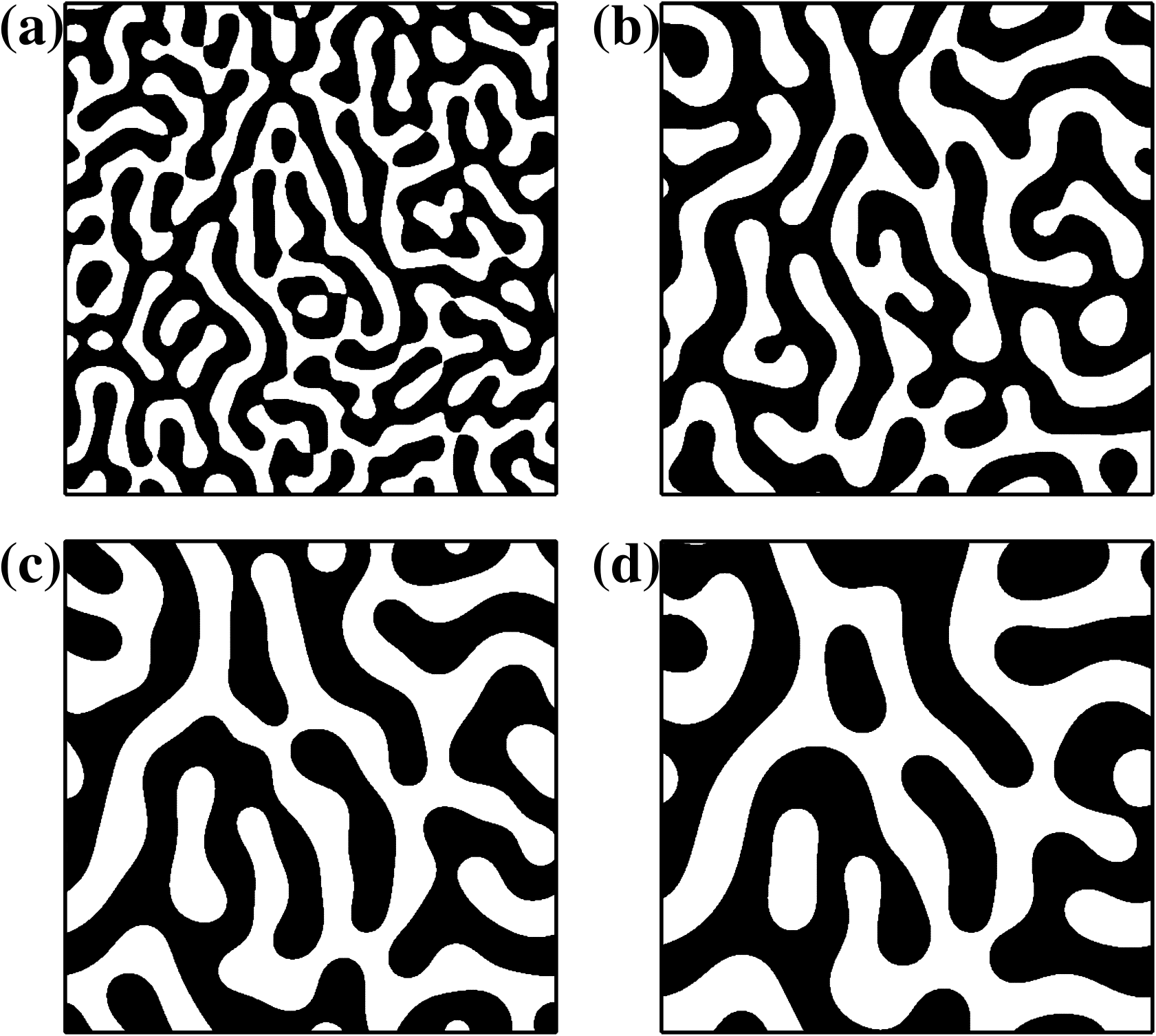}
\caption{\label{fig1} Evolution snapshots of a symmetric active binary mixture (50\%A - 50\%B) at different times, (a) $t=500$, (b) $t=2000$, (c) $t=7000$, and (d) $t=10000$ for $\epsilon=0.02$ and $\Delta=1$. Regions with $\psi>0$ are marked in black, while regions with $\psi<0$ are kept unmarked.}
\end{figure}

We begin with the initial densities of species, $\rho_A$ and $\rho_B$ as small fluctuation around the mean density $\rho_0$, i.e., $\rho_i=\rho_0\pm 0.01$ with $\rho_0=1.065$. This corresponds to a homogeneous mixture at $t=0$. For phase separation, we use $+$ signs in eq.~(\ref{apeqn101}) and our choice of densities meets the instability condition as specified in eq.~(\ref{apeqn111}). Figure~\ref{fig1} shows the evolution snapshots of the order parameter field at different times for $\epsilon=0.02$ and $\Delta=1$, as mentioned. At early times, the average domain size is small, and that increases as time advances. In this case, we observed an interconnected or bicontinuous domain structure similar to the Model B~\cite{DP04,PW09}. But, this is in sharp contrast with the AMB ($\alpha\ne 0$), which shows droplet morphology even for symmetric mixtures~\cite{SSS21,SSS2}. Next, we vary $\Delta$ and $\epsilon$ to determine whether these parameters have any effect on the domain structures.

\begin{figure}
\centering
\includegraphics*[width=0.95\textwidth]{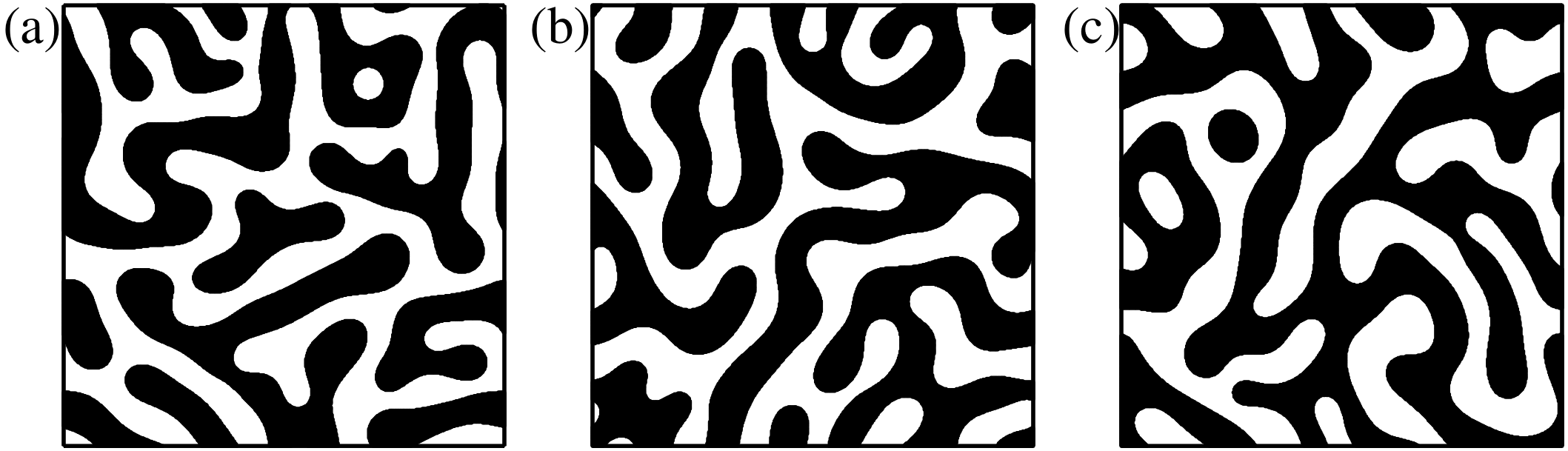}
\caption{\label{fig1a} Evolution snapshots of a symmetric active binary mixture (50\%A - 50\%B) at $t=5000$ for $\epsilon=0.04$ and varying relative activity strengths $\Delta$: (a) $\Delta=0.8$, (b) $\Delta=1$,  and (c) $\Delta=2$. Other simulation details are same as in fig~\ref{fig1}.}
\end{figure}

Figure~\ref{fig1a} shows the evolution snapshots of the $\psi(\vec r, t)$-field at $t=5000$ for $\epsilon=0.04$ and different values of $\Delta$, as mentioned. By inspection, we can conclude that the domain structure consistently remains bicontinuous in all cases. This contrasts with the AMB, where changes in $\alpha$ lead to significant visual alterations in the domain structures.~\cite{SSS21,SSS2}. Next, we vary $\epsilon$ while keeping $\Delta$ constant and observe that the bicontinuous domain structure, similar to that shown in Fig.~\ref{fig1a}, persists across different values of $\epsilon$ (results not shown).

\begin{figure}
\centering
\includegraphics*[width=0.90\textwidth]{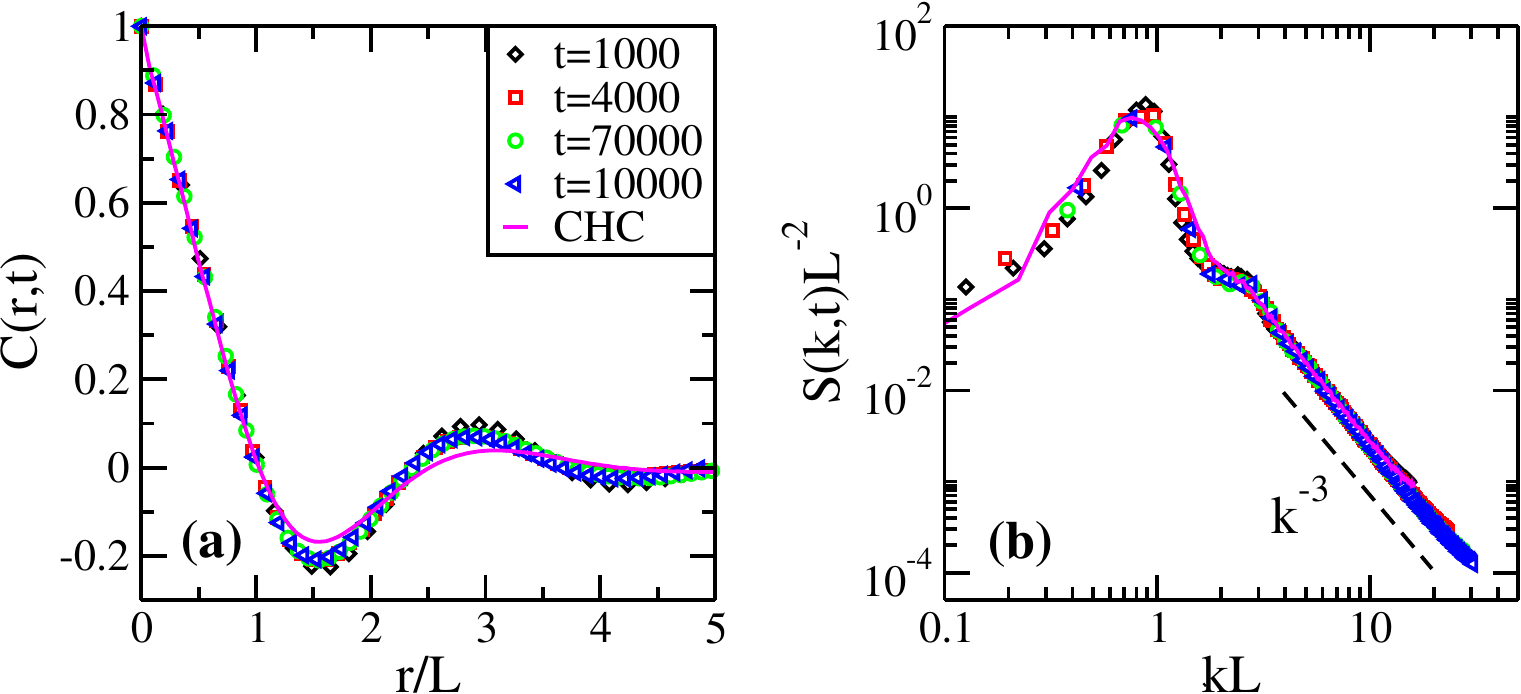}
\caption{\label{fig2} Scaling plot of correlation functions and structure factors for the evolution shown in fig.~\ref{fig1}. (a) Plot of $C(r, t)$ vs. $r/L(t)$ at indicated times. We define the length scale $L(t)$ as the first zero crossing of $C(r,t)$. The solid magenta line represents the scaled correlation function $g_p[r/L(t)]$ of Model B (CHC equation) for symmetric binary mixtures. (b) Log-log plot of $S(k, t)L^{-2}$ vs. $kL(t)$ at different times. The symbols used have the same meaning as in (a). The dashed line labeled $k^{-3}$ shows \textit{Porod's law} for $d=2$.}
\end{figure}

Next, we characterize the morphology of domain growth by computing the equal-time correlation function $C(\vec r, t)$ and structure factor $S(\vec k, t)$. For the order parameter $\psi(\vec r, t)$, the equal-time correlation function is defined as 
\begin{eqnarray}
\label{apeqn12}
C(r, t)=\bigl\langle\psi(\vec{R}+\vec{r}, t)\psi(\vec{R}, t)\bigr\rangle-\bigl\langle\psi(\vec{R}+\vec{r}, t) \bigr \rangle\bigl\langle\psi(\vec{R}, t)\bigr\rangle.
 \label{sec3}
\end{eqnarray}
Here the angular brackets denote an average over reference positions $\vec{R}$ and spherical averaging over different directions. The structure factor $S(\vec k, t)$ is defined as the Fourier transform of $C(\vec r, t)$ at wave vector $\vec k$ as 
\begin{eqnarray}
\label{apeqn13}
S(\vec k, t)  = \int d\Vec{r}  e^{i\Vec{k}.\Vec{r}} C(\Vec{r},t).
\end{eqnarray}
In fig.~\ref{fig2}(a), we have plotted $C(r, t)$ vs. $r/L(t)$ for $\Delta=1$ and $\epsilon=0.02$ at different times, as mentioned. Here $L(t)$ is the average domain size. Clearly, numerical data at different times collapse on each other, which indicates dynamical scaling. However, the scaling function is different from the master function $g_p[r/L(t)]$ of the Model B for symmetric mixtures at large distances, as indicated by the solid magenta line. In fig.~\ref{fig2}(b), we have shown a plot of  $S(k, t)L^{-2}$ vs. $kL(t)$ at different times on a log-log scale. Similar to $C(r, t)$, $S(k, t)$ also shows dynamical scaling. In the limit $k\rightarrow\infty$, the tail of $S(k, t)$ decays as $k^{-3}$, following the well-known \textit{Porod's Law}: $S(k, t)\sim k^{-(d+1)}$ in $d=2$, similar to Model B. This occurs due to scattering from sharp interfaces between A-rich and B-rich domains~\cite{GP82,OP88}. The occurrence of dynamical scaling and Porod's law holds true for other values of $\Delta$ as well.

\begin{figure}
\centering
\includegraphics*[width=0.95\textwidth]{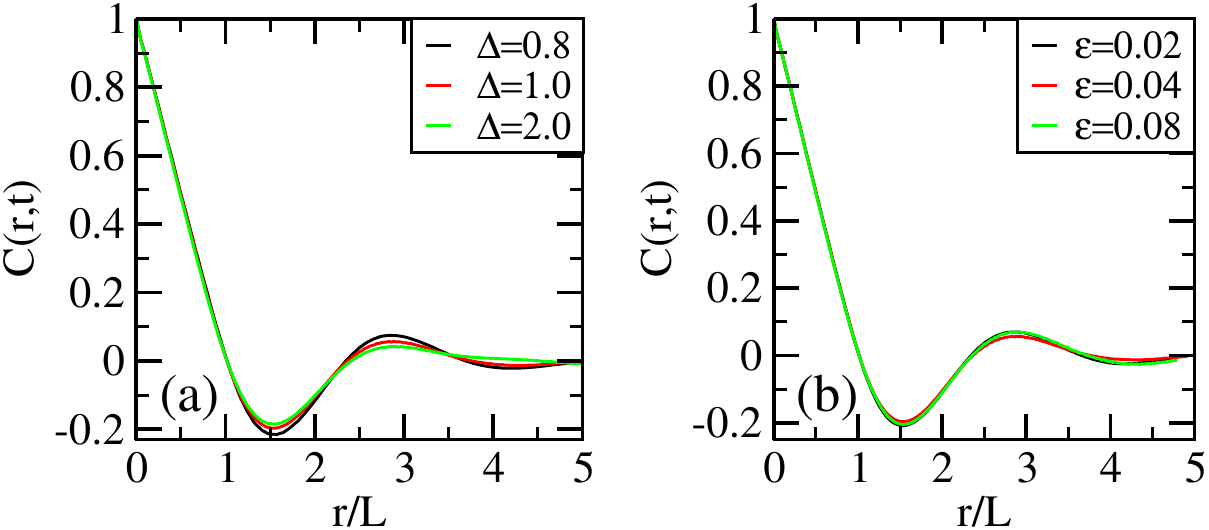}
\caption{\label{fig2a} Scaled correlation functions $C(r, t)$ for symmetric mixtures at $t=5000$. (a) Plot of $C(r, t)$ vs. $r/L(t)$ for $\epsilon=$ 0.04 and different values of $\Delta$, as mentioned. (b) Plot of $C(r, t)$ vs. $r/L(t)$ for $\Delta=1$ and different values of $\epsilon$, as indicated.}
\end{figure}

Next, to reveal the effect of $\Delta$ and $\epsilon$ on domain morphologies, we plot the scaled $C(r, t)$ at $t=5000$ for various values of $\Delta$ and $\epsilon$ in fig~\ref{fig2a}. Figure~\ref{fig2a}(a) shows plot of $C(r, t)$ vs. $r/L(t)$ for $\epsilon=0.04$ and different values of $\Delta$. The numerical data for various $\Delta$ values are clearly distinguishable at large distances, indicating that the scaled correlations depend on the parameter $\Delta$. This may be attributed to the difference in diffusivities of the species when $\Delta\ne 1$. Figure~\ref{fig2a}(b) shows the plot of $C(r, t)$ vs. $r/L(t)$ for $\Delta=1$ and different values of $\epsilon$. The numerical data for different $\epsilon$ values clearly overlap, indicating that noise has no significant impact on the domain morphologies in the scaling regime, similar to what is observed in Model B and AMB. Increasing noise strength primarily enhances the roughness of domain interfaces, resulting in a delayed transition to the asymptotic scaling regime. If $w$ is the thickness of the domain interface,  the system will reach the scaling regime when $w/L(t)\rightarrow 0$. Furthermore, we observed that $S(k,t)$ follows Porod's law at large $k$ in the scaling regime for all cases (not shown here).

\begin{figure}
\centering
\includegraphics*[width=0.95\textwidth]{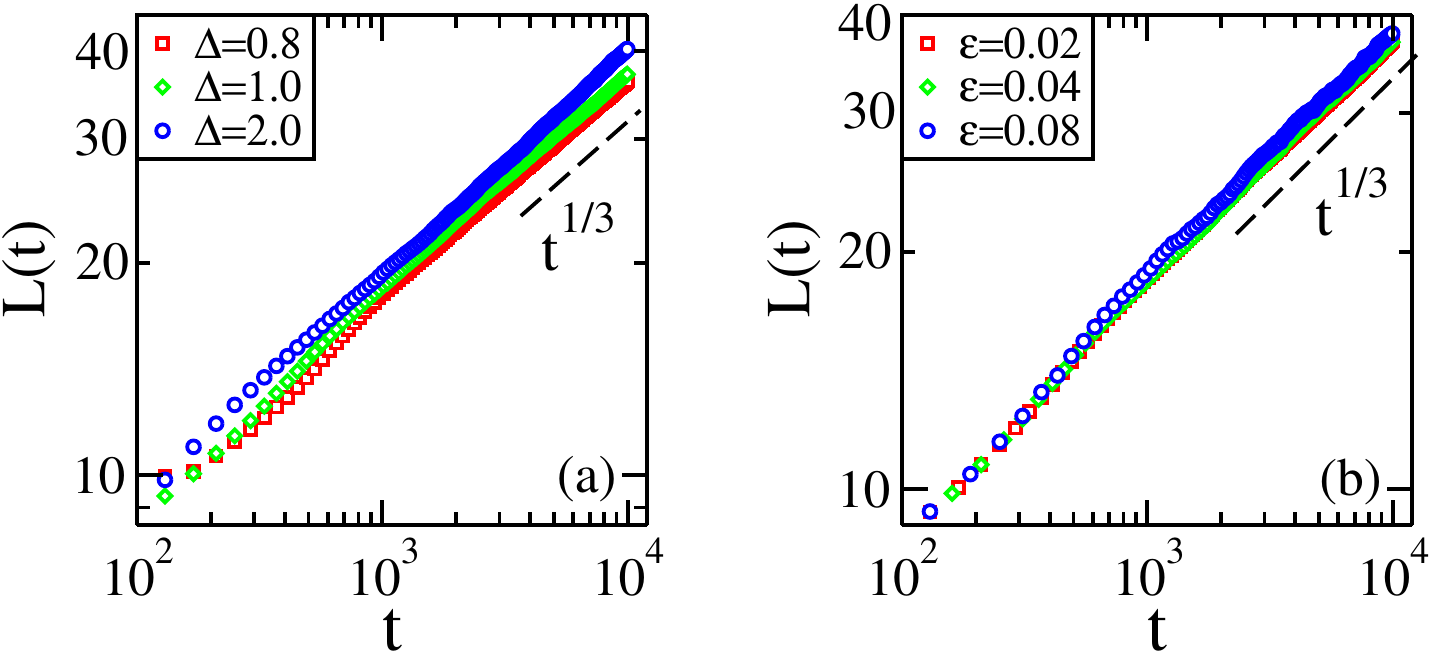}
\caption{\label{fig3} Time dependence of average domain size $L(t)$ for symmetric active binary mixtures. (a) Plot of $L(t)$ vs. $t$ on a log-log scale for $\epsilon=0.02$ and different values of $\Delta$, as mentioned. (b) Plot of $L(t)$ vs. $t$ on a log-log scale for $\Delta=1$ and various values of $\epsilon$, as specified. The dashed black line labeled $t^{1/3}$ represents the Lifshitz-Slyozov (LS) law: $L(t)\sim t^{1/3}$.}
\end{figure}

Figure~\ref{fig3} shows the plot of the average domain size $L(t)$ vs. $t$ on a log-log scale for different values $\Delta$ and $\epsilon$. We define $L(t )$ as the distance over which the correlation function $C(r,t)$ decays to zero for the first time from its maximum value at $r=0$. In fig~\ref{fig3}(a), we have plotted $L(t)$ vs. $t$ for $\epsilon=0.04$ and different values of $\Delta$. It is evident that in the late stage of evolution, $L(t)$ follows a power law growth, $L(t)=A_a t^{1/3}$, consistent with the LS law, similar to what is observed in Model B. The prefactor $A_a$ increases with $\Delta$ because the diffusivity of the B component rises as $\Delta$ increases. Figure~\ref{fig3}(b) shows $L(t)$ vs. $t$ for $\Delta=1$ and various values of $\epsilon$. Once again, $L(t)$ follows the LS law. However, the prefactor $A_a$ only weakly dependent on $\epsilon$. These results confirm that bulk diffusion drives the phase separation in symmetric active binary mixtures.

\subsection{Asymmetric Case}\label{sec3b}
\begin{figure}
\centering
\includegraphics*[width=0.55\textwidth]{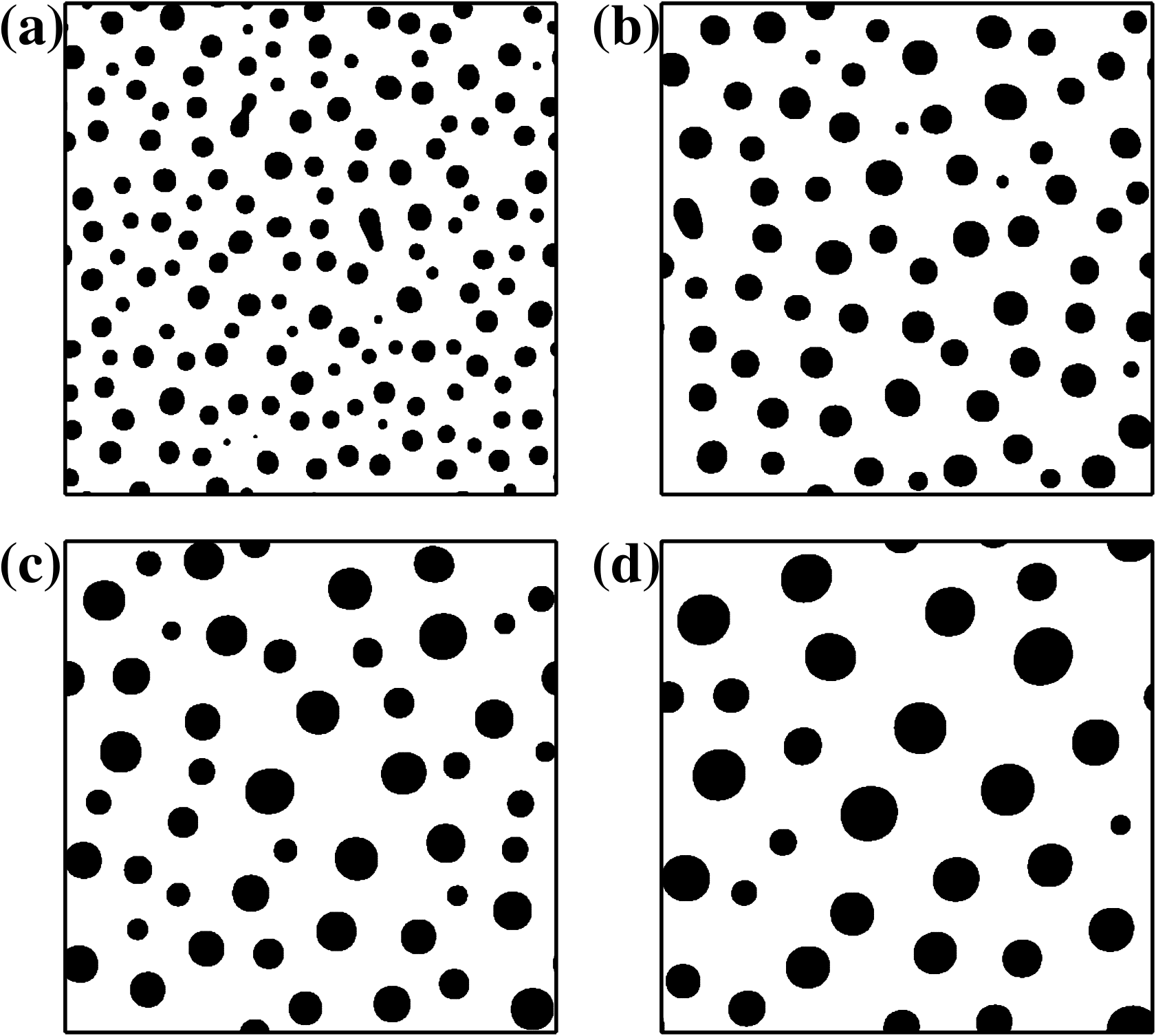}
\caption{\label{fig4} Evolution snapshots of an asymmetric active binary mixture (30\%A - 70\%B) at different times, (a) $t=500$, (b) $t=2000$, (c) $t=7000$, and (d) $t=10000$ for $\epsilon=0.02$ and $\Delta=1$. A-rich regions with $\psi>0$ are marked in black, while B-rich regions with $\psi<0$ are kept unmarked.}
\end{figure}

First we discuss kinetics of phase separation in 30\%A-70\%B asymmetric active binary mixtures in details. We begin with a homogeneous mixture with initial density of species A as $\rho_A=0.695\pm 0.01$ and for species B as $\rho_B=1.625\pm 0.01$ at $t=0$. Clearly, our choice of densities satisfies the instability condition specified in eq.~(\ref{apeqn111}), and we use the $+$ signs in eq.~(\ref{apeqn101}) for phase separation. Figure~\ref{fig4} shows the evolution snapshots of the $\psi(\vec r,t)$-field at different times, as mentioned. We observe circular droplet-like morphology similar to the off-critical Model B and AMB. At the early stage of evolution, the average droplet size is small, and that increases as time increases. 

\begin{figure}
\centering
\includegraphics*[width=0.99\textwidth]{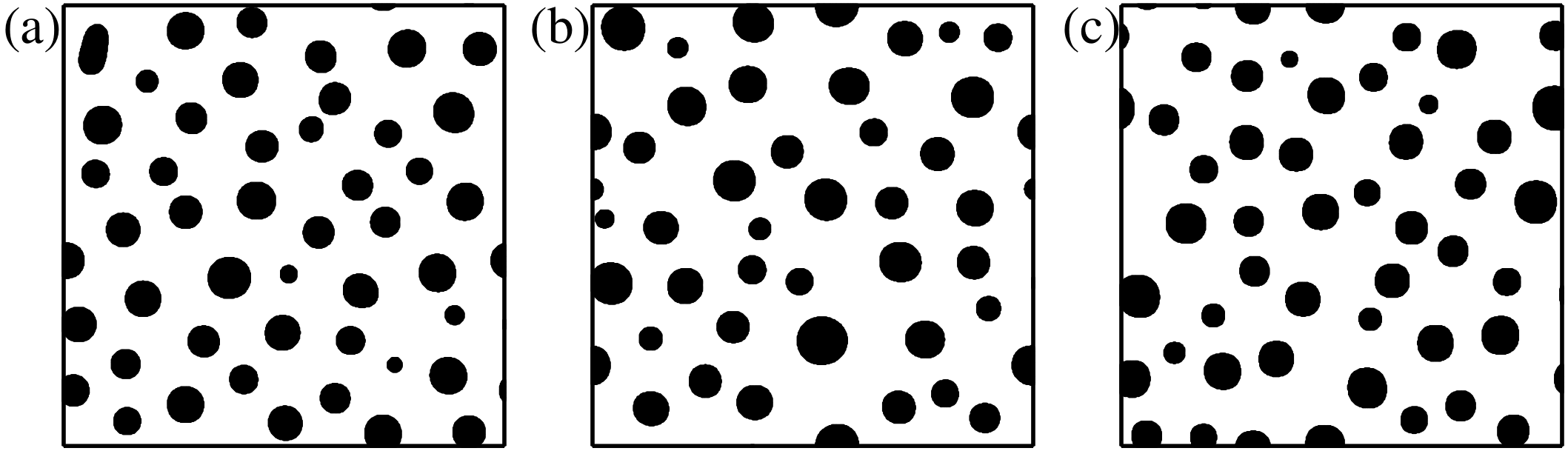}
\caption{\label{fig5a} Evolution snapshots of an asymmetric active binary mixture (30\%A - 70\%B) at $t=5000$ for $\epsilon=0.04$ and varying relative activity strengths $\Delta$: (a) $\Delta=0.8$, (b) $\Delta=1$,  and (c) $\Delta=2$. Other simulation details are same as in fig~\ref{fig4}.}
\end{figure}

Next, we examine the impact of relative activity parameter $\Delta$ and noise $\epsilon$ on the domain morphology. Figure~\ref{fig5a} presents snapshots of the $\psi(\vec r, t)$-field at $t=5000$ for $\epsilon=0.04$ and various values of $\Delta$, as mentioned. Similar to fig.~\ref{fig4}, we observe circular domains of minority components, regardless of $\Delta$ values. Additionally, we varied $\epsilon$ for a fixed $\Delta$ and observed that the droplet domain morphology of the minority component remains consistent across different $\epsilon$ values (results no shown), with no noticeable difference from fig.~\ref{fig5a}.

\begin{figure}
\centering
\includegraphics*[width=0.95\textwidth]{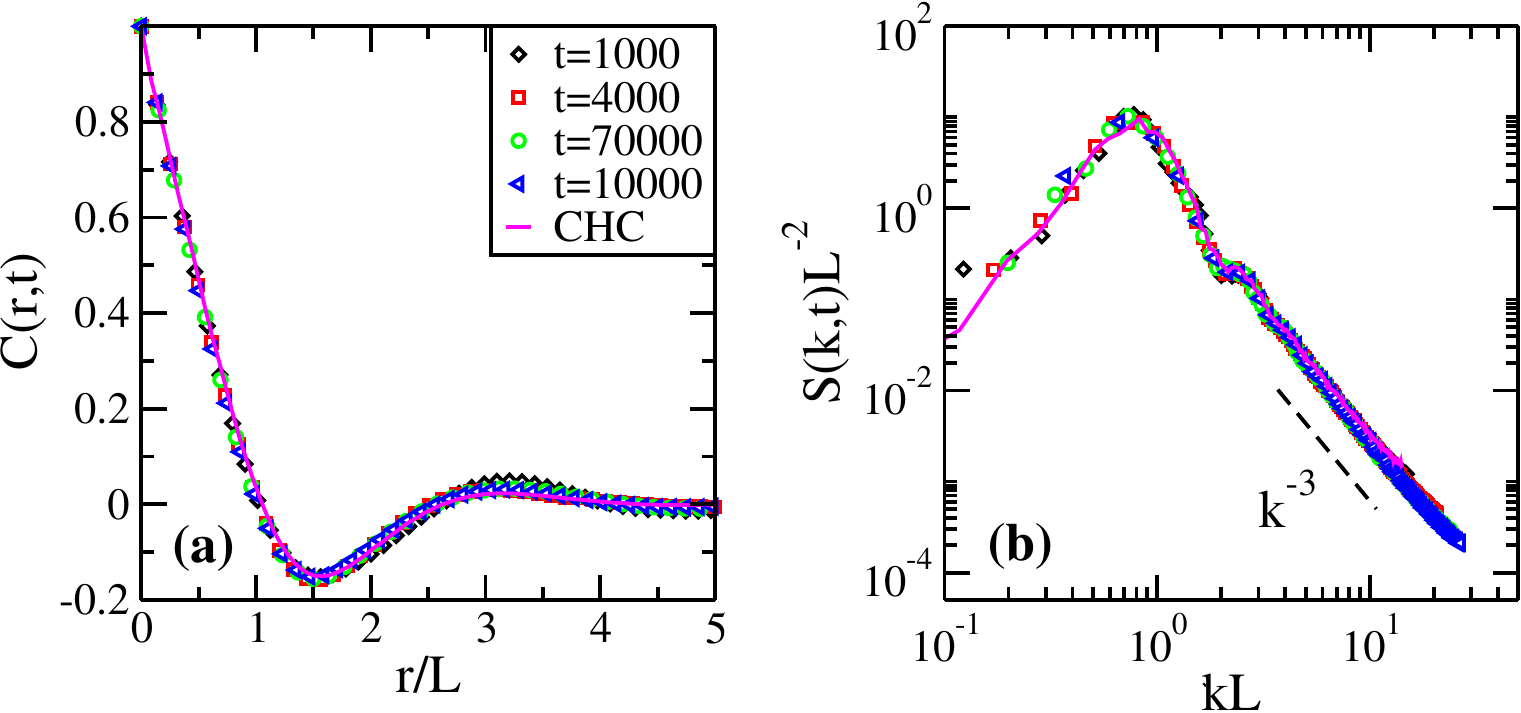}
\caption{\label{fig5} Scaling plot of equal-time correlation functions $C(r,t)$ and structure factors $S(k,t)$ for an asymmetric active binary mixture (30\%A - 70\%B) for $\Delta=1$ and $\epsilon=0.02$. (a) Plot of correlation function $C(r, t)$ vs. $r/L(t)$ at different times, as mentioned. The solid magenta line represents the scaled correlation function $g_p[r/L(t)]$ from Model B for the same mixture composition. (b) Plot of structure factor $S(k, t)L^{-d}$ vs. $kL(t)$ on a log-log scale at different times, as indicated. The dashed line labeled $k^{-3}$ shows \textit{Porod's law} for $d=2$.}
\end{figure}

To characterize the morphology of domains, we again compute $C(r, t)$ and $S(k, t)$ as given by eqs.~(\ref{apeqn12}) and (\ref{apeqn13}) respectively. Figure~\ref{fig5}(a) shows the plot of $C(r, t)$ vs. $r/L(t)$ for $\Delta=1$ and $\epsilon=0.02$ at different instants, as mentioned. Similar to the symmetric case, numerical data at different times collapse on top of each other. This confirms dynamical scaling. Moreover, unlike the symmetric case, the scaled correlation function is almost similar to $g_p[r/L(t)]$ from Model B for the same mixture composition, as represented by the solid magenta line. In fig.~\ref{fig5}(b), we have plotted structure factor $S(k, t)L^{-2}$ vs. $kL(t)$ on a log-log scale for different times. Similar to $C(r, t)$, $S(k, t)$ also shows dynamical scaling. In the limit $k\rightarrow\infty$, the tail of $S(k, t)$'s also follow the \textit{Porod's Law}. These results hold true for other values of $\Delta$ as well. 

\begin{figure}
\centering
\includegraphics*[width=0.95\textwidth]{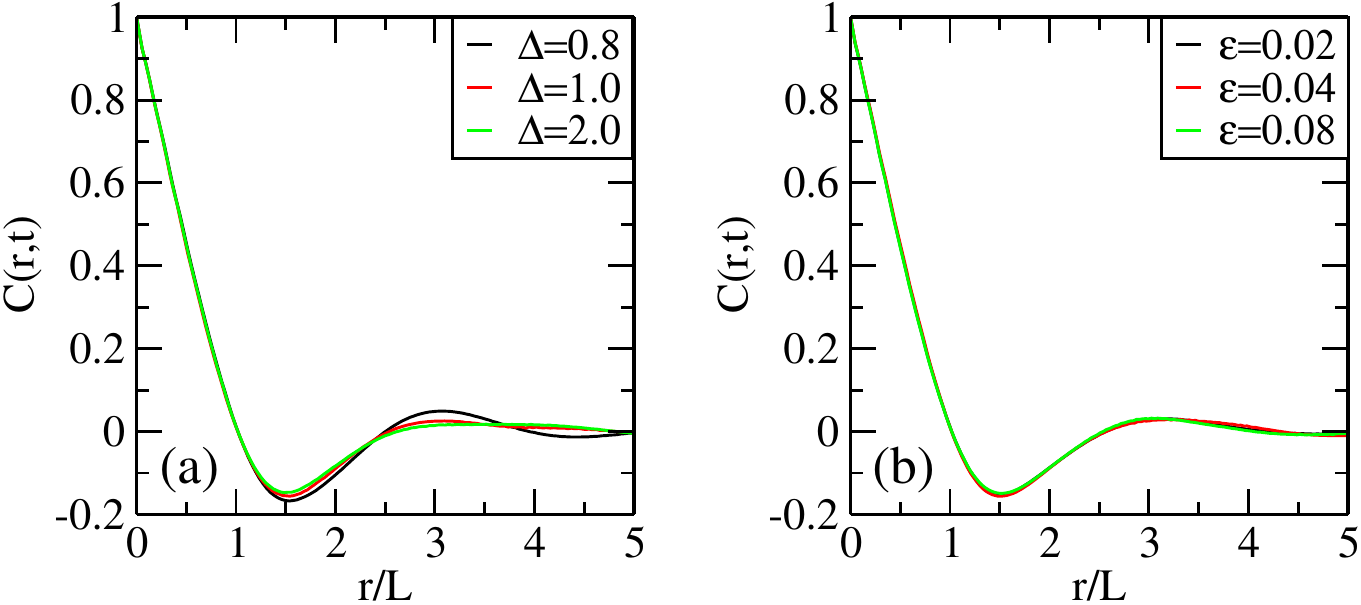}
\caption{\label{fig6a} Scaled correlation functions $C(r, t)$ for 30\%A - 70\%B asymmetric mixtures at $t=5000$. (a) Plot of $C(r, t)$ vs. $r/L(t)$ for $\epsilon=$ 0.04 and different values of $\Delta$, as mentioned. (b) Plot of $C(r, t)$ vs. $r/L(t)$ for $\Delta=1$ and different values of $\epsilon$, as indicated.}
\end{figure}

Next, we compute the scaled $C(r, t)$ at $t=5000$ for various values of $\Delta$ and $\epsilon$. Figure~\ref{fig6a}(a) shows the plot of $C(r, t)$ vs. $r/L(t)$ for $\epsilon=0.04$ and different values of $\Delta$. The numerical data for the different $\Delta$ values are clearly distinguishable at large distances, with the depth of the first minima of $C(r, t)$ systematically varying as $\Delta$ changes. This confirms that the scaled correlations depend on the parameter $\Delta$, similar to what is observed in symmetric mixtures. Figure~\ref{fig6a}(b) shows the plot of $C(r, t)$ vs. $r/L(t)$ for $\Delta=1$ and different values of $\epsilon$. The numerical data for different $\epsilon$ values clearly overlap, confirming that noise has no significant impact on the domain morphologies in the scaling regime, similar to what is observed in symmetric mixtures. Again, we observed that $S(k,t)$ follows Porod's law at large $k$ for all cases (not shown here).

\begin{figure}
\centering
\includegraphics*[width=0.95\textwidth]{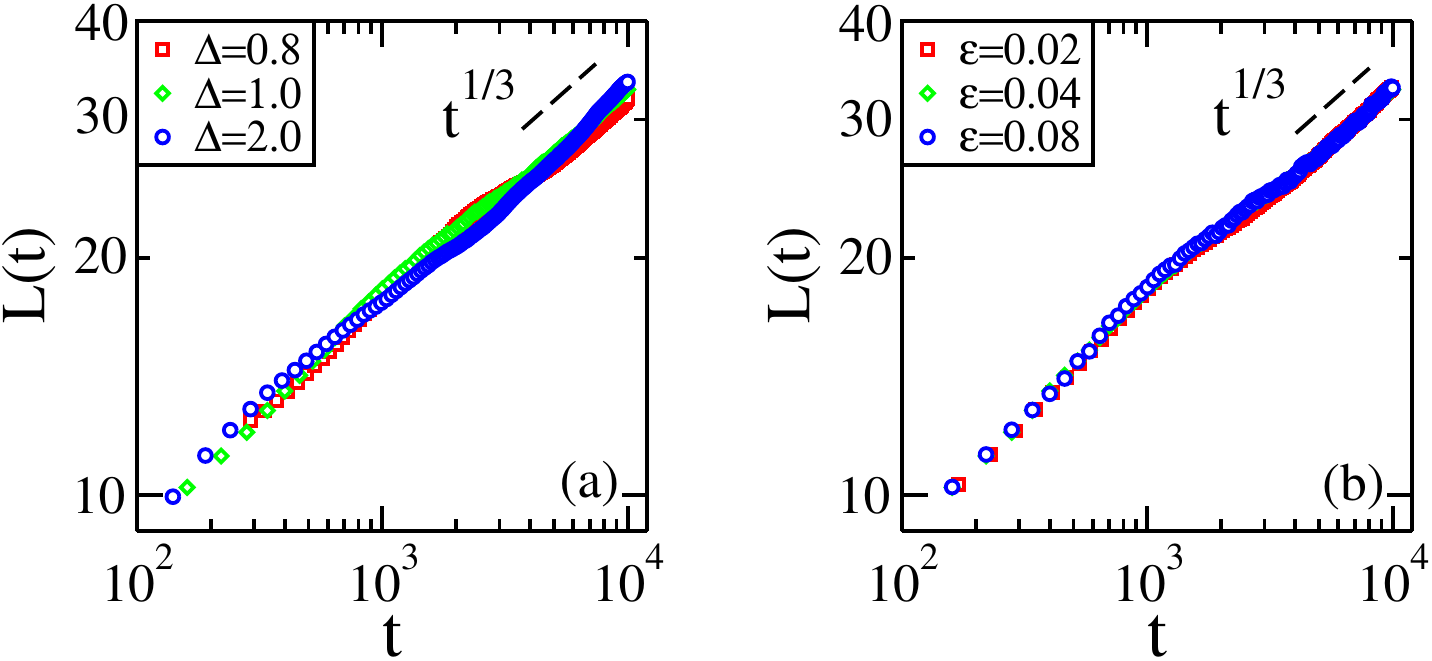}
\caption{\label{fig6} Figure analogous to Fig.~\ref{fig3} for an asymmetric active binary mixture (30\%A - 70\%B). (a) Plot of $L(t)$ vs. $t$ on a log-log scale for $\epsilon=0.04$ and different values of $\Delta$, as mentioned. (b) Plot of $L(t)$ vs. $t$ on a log-log scale for $\Delta=1$ and various values of $\epsilon$, as specified. The black dashed line represents the LS law: $L(t)\sim t^{1/3}$.}
\end{figure}

We calculate the average domain size $L(t)$ from the decay of the correlation function $C(r, t)$. Figure~\ref{fig6} shows plot of $L(t)$ vs. $t$ on a log-log scale for various values of $\Delta$ and $\epsilon$. It is evident that $L(t)$ follows the LS Law: $L(t)\sim t^{1/3}$ in the large-$t$ regime across all cases, similar to the symmetric case. Although the growth exponent is the same as in Model B kinetics,  a key difference is that, unlike in Model B, where the diffusivity $D_p$ is considered to be a constant, here the diffusivities of the species in our model depend on densities $\rho$.

\begin{figure}
\centering
\includegraphics*[width=0.95\textwidth]{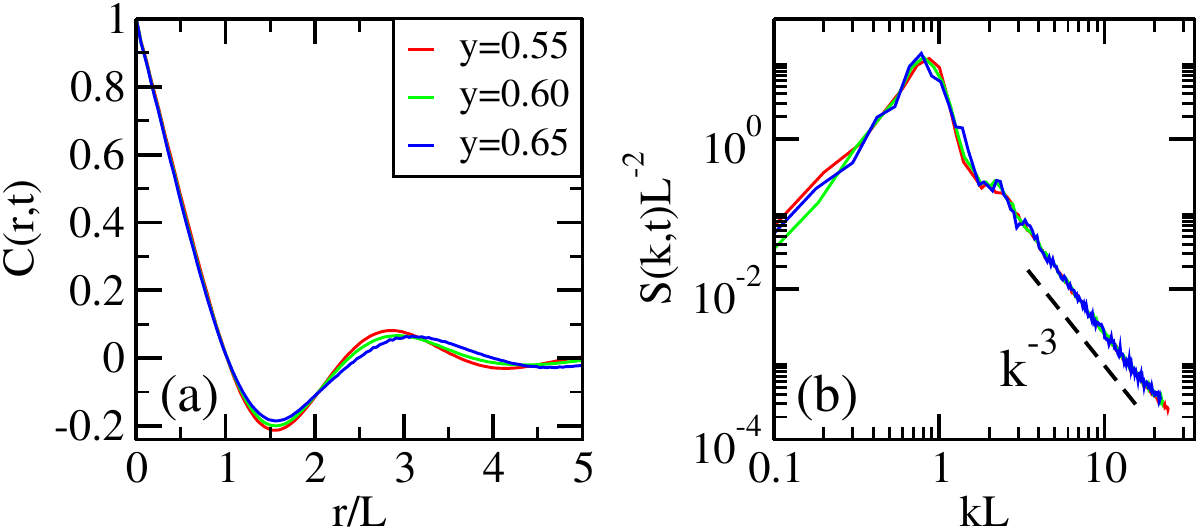}
\caption{\label{fig11} Scaled correlation functions $C(r, t)$ and structure factors $S(k, t)$ at $t=5000$ for various asymmetric mixtures. We set $\Delta=1$ and $\epsilon=0.02$. (a) Plot of $C(r, t)$ vs. $r/L(t)$  for different values of $y$, as mentioned. (b) Plot of $S(k, t)$ vs. $kL(t)$ on a log-log scale. The symbols used have the same meaning as in (a). The dashed blacked line, labeled $k^{-3}$, represents Porod's law.}
\end{figure}

Finally, we focus on how domain morphologies and growth laws depend on the mixture compositions. We simulate several other asymmetric mixtures with $y$ values ranging from 0.55 to 0.65. Table~\ref{Table_1} lists the initial homogeneous densities at $t=0$, which satisfy the instability condition defined by eq.~(\ref{apeqn111}). In all cases, the systems evolve by forming droplet morphologies of minority components, similar to the 30\%A-70\%B mixture shown in fig.~\ref{fig4}. We observed that for a given $y$ and $\Delta$, both $C(r,t)$ and $S(k,t)$ follow dynamical scaling, with $S(k,t)$ follows Porod's law at large wave vectors. Additionally,, for a given $y$, the scaled $C(r,t)$ depends on $\Delta$ but is independent of $\epsilon$, consistent with the behavior seen in both symmetric and asymmetric mixtures discussed earlier. Also, we compared the $C(r,t)$ obtained from the current model for $\Delta=1$ with that of from Model B and observed deviation between the two for a given $y$. The degree of deviation depends on the value of $y$. The average domain size $L(t)$ follows the LS law, regardless of the values of $y$, $\Delta$, and $\epsilon$. These parameters only influence the prefactor of the growth law. Now, a pertinent question remains: do the scaling functions for a fixed $\Delta$ depend explicitly on $y$? We present scaled $C(r, t)$ and $S(k, t)$ at $t=5000$ for $\Delta=1$, $\epsilon=0.02$ and various values of $y$ in fig.~\ref{fig11}. $C(r, t)$ for different $y$ values systematically deviates at large distances as $y$ changes as shown in fig.~\ref{fig11}(a), indicating that the scaled $C(r,t)$ depends on the the system's composition. This behavior is consistent with both Model B and AMB~\cite{SP88,SSS21,SSS2,ASSP15}. Next, we plot $S(k, t)$ vs. $kL(t)$ on a log-log scale in fig.~\ref{fig11}(b). We observe a deviation of $S(k,t)$ in the small-$k$ regime, corresponding to the deviation seen in $C(r,t)$ at large distances. However, as $k\rightarrow\infty$, $S(k, t)$ follows Porod's law, $S(k, t)\sim k^{-3}$ in $d=2$, regardless of the $y$ values. This behavior is again consistent with both Model B and AMB~\cite{SP88,SSS21,SSS2,ASSP15}.

\section{Summary and Outlook}\label{sec4}
Let us conclude this paper with a summary and discussion of our results. We have studied motility-induced phase separation (MIPS) in A$_{1-y}$B$_y$ active binary mixtures using a coarse-grained model. The two-dimensional run-and-tumble of bacterial motion underpins microscopic dynamics. The model incorporates the relative activity of the species $\Delta$ and noise strength $\epsilon$. A mixture is said to be symmetric when $y=0.5$ and asymmetric otherwise. At $t=0$, we begin with a homogeneous AB mixture, selecting the densities $\rho_i$ to ensure that the instability condition is satisfied. As time passes, the system phase separates into A-rich and B-rich domains. A symmetric mixture shows an interconnected or bi-continuous domain structure, regardless of the values of $\Delta$ and $\epsilon$, similar to Model B. In contrast, the AMB can display a droplet-like morphology, even in symmetric mixtures. For asymmetric mixtures, however, all models show a circular droplet morphology of the minority component.

We have characterized the domain growth morphology by computing the equal-time correlation function $C(r,t)$ and structure factor $S(k,t)$. For a given mixture composition (fixed $y$) and $\Delta$, the scaled correlation functions and structure factors follow dynamical scaling, with the form of the scaling functions being independent of $\epsilon$. However, for a given mixture composition, the form of the scaling functions depends on $\Delta$, and for a fixed $\Delta$, it depends on the mixture composition, regardless of $\epsilon$. Furthermore, these scaling functions for $\Delta=1$ differ from those of Model B for all mixture compositions, with the largest deviation occurring in symmetric mixtures. In the limit $k \rightarrow \infty$, $S(k, t)$ follows Porod's law for all mixtures, irrespective of the values of $\Delta$ and $\epsilon$. We have estimated the average domain size $L(t)$ from the decay of $C(r, t)$. $L(t)$ follows the Lifshitz-Slyozov growth law: $L(t) \sim t^{1/3}$ for all compositions, similar to Model B. This is due to diffusion being the mechanism of domain growth. AMB, on the other hand, exhibits a crossover from $L(t) \sim t^{1/3}$ at early times to $L(t) \sim t^{1/4}$ at a later time.

We believe that our current results provide a detailed understanding of MIPS in active mixtures in $d=2$. Understanding complex domain growth kinetics in multicomponent mixtures, where some of the components may be assumed to be passive or dead, and incorporating chemical reaction among the components, would be an intriguing avenue for further research. We also aim to investigate the impact of spatial dimensionality and incorporate hydrodynamics into this model. In passive systems, the effect of hydrodynamics causes a crossover in domain growth exponents, and they depend on the spatial dimension~\cite{EDS79,FURU85,MGG85,DDP24}. Therefore, we expect these problems to offer opportunities for discovering different domain growth exponents, scaling violations, and other rich dynamical behaviors. Additionally, these studies hold significant relevance for various experimental scenarios, including applications in biological physics and engineering.

\ \\
\noindent{\bf Acknowledgments:} We thank Sanjay Puri from JNU for fruitful discussions and are grateful to the referees for their constructive and useful suggestions. SM acknowledges financial support from IISER Mohali through a Junior Research Fellowship. PD acknowledges financial support from SERB, India through a start-up research grant (SRG/2022/000105).

\end{document}